\documentclass[twocolumn,aps,pra,notitlepage,superscriptaddress,longbibliography]{revtex4-1}
\usepackage[colorlinks, linkcolor = ForestGreen, anchorcolor = blue, citecolor = magenta]{hyperref}
\usepackage{amsmath,amssymb,amsthm,mathtools,mathrsfs}
\usepackage[table,dvipsnames]{xcolor}
\usepackage{tikz}
\usepackage{adjustbox}
\usepackage{dsfont}
\usepackage[normalem]{ulem}
\usepackage{enumerate}
\usepackage[all]{xy} 

\theoremstyle{plain}
\newtheorem{theorem}{Theorem}
\newtheorem{lemma}{Lemma}
\newtheorem{proposition}{Proposition}

\theoremstyle{definition}
\newtheorem{definition}{Definition}
\newtheorem{remark}{Remark}
\newtheorem{example}{Example}

\newcommand{\bd}{\begin{definition}}
\newcommand{\ed}{\end{definition}}
\newcommand{\bt}{\begin{theorem}}
\newcommand{\et}{\end{theorem}}
\newcommand{\bn}{\begin{proposition}}
\newcommand{\en}{\end{proposition}}
\newcommand{\be}{\begin{equation}}
\newcommand{\ee}{\end{equation}}
\newcommand{\blem}{\begin{lemma}}
\newcommand{\elem}{\end{lemma}}
\newcommand{\bx}{\begin{example}}
\newcommand{\ex}{\end{example}}
\newcommand{\bprf}{\begin{proof}}
\newcommand{\eprf}{\end{proof}}

\newcommand\define[1]{\emph{\textbf{#1}}}

\DeclareMathAlphabet{\mathpzc}{OT1}{pzc}{m}{it} 
 \DeclareFontFamily{OT1}{pzc}{}
 \DeclareFontShape{OT1}{pzc}{m}{it}{ <-> s*[1.2] pzcmi7t }{}
 \DeclareMathAlphabet{\mathpzc}{OT1}{pzc}{m}{it}
 \newcommand{\Alg}[1]{\mathpzc{#1}}

\newcommand{\<}{\langle}
\renewcommand{\>}{\rangle}

\newcommand{\Tr}{\operatorname{Tr}}

\def\A{\Alg{A}}
\def\B{\Alg{B}}

\def\E{\mathcal{E}}
\def\H{\mathcal{H}}
\def\F{\mathcal{F}}
\def\O{\mathscr{O}}

\def\J{\mathscr{J}}

\def\S{\mathfrak{S}}

\begin{document}									
\preprint{APS/123-QED}

\title{The spatiotemporal Born rule is quasiprobabilistic}

\author{James Fullwood}
\affiliation{School of Mathematics and Statistics, Hainan University, Haikou, Hainan Province, 570228, China}
\author{Zhihao Ma}
\affiliation{School of Mathematical Sciences, MOE-LSC, Shanghai Jiao Tong University, Shanghai, 200240, China}
\affiliation{Shanghai Seres Information Technology Co., Ltd, Shanghai 200040, China}
\affiliation{Shenzhen Institute for Quantum Science and Engineering, Southern University of Science and Technology, Shenzhen 518055, China}
\author{Zhen Wu}
\email{wzmath@hainanu.edu.cn}
\affiliation{School of Mathematics and Statistics, Hainan University, Haikou, Hainan Province, 570228, China}
\affiliation{School of Mathematical Sciences, MOE-LSC, Shanghai Jiao Tong University, Shanghai, 200240, China}

\date{\today}

\begin{abstract}
 Contrary to general relativity, quantum theory treats space and time in fundamentally different ways. In particular, while joint probabilities associated with spacelike separated measurements are defined in terms of the Born rule, joint probabilities associated with measurements performed in sequence are defined in terms of the state-update rule. In this work, we show that one obtains a more unified perspective of space and time in quantum theory by embracing a quasiprobabilistic description of sequential measurements. More precisely, we show that there exists a unique \emph{pseudo}-density operator encoding canonical quasiprobabilities associated with sequential measurements in precisely the same manner that a density operator encodes joint probabilities associated with spacelike separated measurements, thus providing a natural extension of the Born rule into the temporal domain. As an application, we show how such a spatiotemporal Born rule combined in conjunction with a quantum Bayes' rule yields an operational notion of time-reversal symmetry for sequential measurements on an \emph{open} quantum system. 
\end{abstract}

	\maketitle

\section{Introduction}

Let $A$ and $B$ denote finite-dimensional quantum systems, and let $\{P_i\}$ and $\{Q_j\}$ denote projective measurements on $A$ and $B$, respectively. If $A$ and $B$ are spacelike separated, then the measurements $\{P_i\}$ and $\{Q_j\}$ may be performed in parallel, and in such a case the Born rule tells us that the joint probability $\bold{P}_{\text{space}}(i,j)$ of measurement outcomes $P_i$ and $Q_j$ is given by
\be \label{PAXRQ87}
\bold{P}_{\text{space}}(i,j)=\Tr[\rho_{AB}(P_i\otimes Q_j)]\, ,
\ee
where $\rho_{AB}$ is the density operator representing the joint state of the composite system $AB$ prior to measurement. Moreover, since the Born rule \eqref{PAXRQ87} holds as $\{P_i\}$ and $\{Q_j\}$ vary across all possible projective measurements, the density operator $\rho_{AB}$ defines a bi-additive measure $\mu$ on the entire space of separable projectors given by 
\[
\mu(P\otimes Q)=\Tr[\rho_{AB}(P\otimes Q)]\, ,
\]
so that $\mu$ satisfies the following properties:
\begin{itemize}
\item
(Normalization) $\mu(\mathds{1}_A\otimes \mathds{1}_B)=1$.
\item
(Local Additivity on $A$) If $\{R_k\}$ is a mutually orthogonal collection of projectors on system $A$, then for every projector $Q$ on system $B$ we have $\mu\Big(\sum_k R_k\otimes Q\Big)=\sum_k\mu(R_k\otimes Q)$.
\item
(Local Additivity on $B$) If $\{S_l\}$ is a mutually orthogonal collection of projectors on system $B$, then for every projector $P$ on system $A$ we have $\mu\Big(P\otimes \sum_lS_l\Big)=\sum_l\mu(P\otimes S_l)$.
\end{itemize}

On the other hand, if the systems $A$ and $B$ are instead timelike separated, so that $\{P_i\}$ and $\{Q_j\}$ may be measured in sequence, then according to the L\"{u}ders-von~Neumann projection postulate \cite{Lu06,vN18}, the probability $\bold{P}_{\text{time}}(i,j)$ of the measurement outcome $P_i$ followed by $Q_j$ is given by
\be \label{PSEXQ87}
\bold{P}_{\text{time}}(i,j)=\Tr[\E(P_i\rho_A P_i) Q_j]\, ,
\ee
where $\rho_A$ is the initial state of system $A$ prior to measurement and $\E$ is a quantum channel responsible for the evolution of the system between measurements. Contrary to the probabilities $\bold{P}_{\text{space}}(i,j)$, there \emph{does not} in general exist a fixed operator $\rho_{AB}$ such that the probabilities $\bold{P}_{\text{time}}(i,j)$ are given by a Born rule of the form \eqref{PAXRQ87} as $\{P_i\}$ and $\{Q_j\}$ vary across all projective measurements. This is due to the fact that $\bold{P}_{\text{time}}$ does not in general induce a locally additive measure on the space of separable projectors. In particular, due either to state disturbance caused by the initial measurement or properties of the dynamics $\E$, there will exist orthogonal projectors $P_{i_1}$ and $P_{i_2}$ such that for the coarse-grained projector $P_{i_0}=P_{i_1}+P_{i_2}$ we have 
\[
\bold{P}_{\text{time}}(i_0,j)\neq \bold{P}_{\text{time}}(i_1,j)+\bold{P}_{\text{time}}(i_2,j)\, ,
\]
thus violating local additivity on $A$. We note that such a disparity between $\bold{P}_{\text{space}}$ and $\bold{P}_{\text{time}}$ is completely at odds with the theory of classical random variables, where properties of joint probabilities are indifferent to whether they are obtained from measurements performed in sequence or in parallel.

The lack of a Born rule of the form \eqref{PAXRQ87} for sequential measurements has thus led to a chasm between space and time in quantum theory. In particular, while the density operator formalism is applied to the study of correlations between spacelike separated systems, the formalism of quantum channels is applied to study of correlations between timelike separated systems (of which unitary evolution is a special case). However, if we are ever to reconcile quantum theory with general relativity, then it will be necessary to develop a single mathematical formalism for quantum theory which treats spatial and temporal correlations in a way which is consistent with the relativistic notion of general covariance, where space and time are treated on equal footing. 

While there are various approaches to reconciling the disparate treatment of space and time in quantum theory~\cite{Chiribella_2008,CDP09,OCB12,LeSp13,CJQW18,Pollock_2018,MaCh22,JiKa23,Huang_2025}, perhaps the most popular approach is that of process matrices, which makes use of quantum instruments to generalize the Born rule in a way that makes no a priori assumption that joint measurements are being performed within a fixed causal structure~\cite{OCB12,Shrapnel_2017}. However, the resulting Born rule is of a different character than that of \eqref{PAXRQ87}. In particular, in the process matrix formalism, the failure of local additivity for the distribution $\bold{P}_{\text{time}}$ is circumvented by defining joint probabilities not on the space of separable projectors, but rather, in terms of a bi-additive measure on the much larger space consisting of separable \emph{instrument components}, whose elements are of the form $\mathcal{M}\otimes \mathcal{N}$, where $\mathcal{M}$ and $\mathcal{N}$ are completely positive, trace non-increasing maps on systems $A$ and $B$, respectively.  


It then follows that the process matrix formalism makes a distinct departure from traditional quantum theory by defining probabilities in a way which differs from the way in which probabilities are defined by the Born rule \eqref{PAXRQ87}. Moreover, while the density operator $\rho_{AB}$ instantiating the Born rule \eqref{PAXRQ87} is a bipartite operator of unit trace, a process matrix associated with joint measurements performed by two agents which are non-spacelike separated is a quadripartite operator of trace greater than unity, which marks yet another way in which the process matrix formalism makes a departure from traditional quantum theory. In light of such observations, here we take the viewpoint that a more natural extension of the Born rule \eqref{PAXRQ87} to the spatiotemporal domain is obtained by embracing a \emph{quasiprobabilistic} description of sequential measurements, which we now explain. 

In Ref.~\cite{Johansen_2007}, Johansen shows that given projective measurements $\{P_i\}$ and $\{Q_j\}$ performed in sequence, there exists a real-valued quasiprobability distribution $\bold{Q}$ such that for all $i$ and $j$,
\be \label{MHX57}
\bold{Q}(i,j)=\bold{P}_{\text{time}}(i,j)+\bold{D}(i,j)\, ,
\ee
where $\bold{D}(i,j)$ is a measure of how the probability of the measurement outcome $Q_j$ is affected by the disturbance of the initial state $\rho_A$ due to an initial measurement outcome of $P_i$. It then follows that any negative values of $\bold{Q}$---which are a signature of non-classicality--- are due to non-zero values of the modification term $\bold{D}$, which occur for example when $[\rho_A,P_i]\neq 0$ or $[P_i,\E^*(Q_j)]\neq 0$, where $\E^*$ is the Hilbert-Schmidt adjoint of the quantum channel $\E$ responsible for the evolution of the system between measurements. Moreover, contrary to $\bold{P}_{\text{time}}$, the quasiprobabilities $\bold{Q}(i,j)$ are bi-additive with respect to local coarse-grainings of measurements, and satisfy the marginal conditions $\sum_j\bold{Q}(i,j)=\Tr[\rho_A P_i]$ and $\sum_i\bold{Q}(i,j)=\Tr[\E(\rho_A)Q_j]$. 

The quasiprobability distribution $\bold{Q}$ is commonly referred to as the \emph{Terletsky-Margenau-Hill} distribution~\cite{Terletsky_1937,Margenau_1961}, which is the real part of a complex-valued quasiprobability distribution introduced independently by Kirkwood and Dirac for the purpose of associating well-defined probabilities with non-commuting observables~\cite{Kirkwood_1933,Dirac_1945}. The study of the Kirkwood-Dirac (KD) and Terletsky-Margenau-Hill (TMH) distributions have undergone something of a renaissance in recent years, especially in the context of quantum thermodynamics, weak values, out-of-time-ordered correlators, quantum metrology and support uncertainty~\cite{BDOV13,Alla_2014,Lost_2018,Alonso_2019,Levy_2020,Lostaglio_2023,Arvidsson_2024,Aharonov_1988,ADYLLBL20,LYABPSH22,Bievre21}. In the context of sequential measurements, formula \eqref{MHX57} yields a precise physical interpretation of the TMH distribution $\bold{Q}$ as the distribution $\bold{P}_{\text{time}}$ corrected by a measure of state disturbance. Moreover, experimental techniques for measuring the quasiprobabilities $\bold{Q}(i,j)$ have recently appeared in the literature~\cite{Hernandez_2024,Wang_2024}, thus establishing an explicit connection between the TMH distribution and experimental physics.

In this work, we show that given a fixed initial state $\rho_A$ and a fixed quantum channel $\E$ governing the dynamics between measurements, then there exists a unique operator $\varrho_{AB}$ such that for all projective measurements $\{P_i\}$ followed by $\{Q_i\}$,
\be \label{STXBORN57}
\bold{Q}(i,j)=\Tr[\varrho_{AB} (P_i\otimes Q_j)]\, .
\ee
As the operator $\varrho_{AB}$ encodes the quasi-probabilities $\bold{Q}(i,j)$ in precisely the same manner that a density operator $\rho_{AB}$ encodes the probabilities $\bold{P}_{\text{space}}(i,j)$, we refer to \eqref{STXBORN57} to as the \emph{spatiotemporal Born rule} for sequential measurements. In light of Eq.~\eqref{MHX57}, the existence of such a spatiotemporal Born rule \eqref{STXBORN57} may then be attributed to the way in which the modification term $\bold{D}(i,j)$ restores local additivity, thus bypassing the effects of back-reaction due to the initial measurement. 

\begin{figure}\label{F1}
    \centering
    \includegraphics[width=0.48\textwidth]{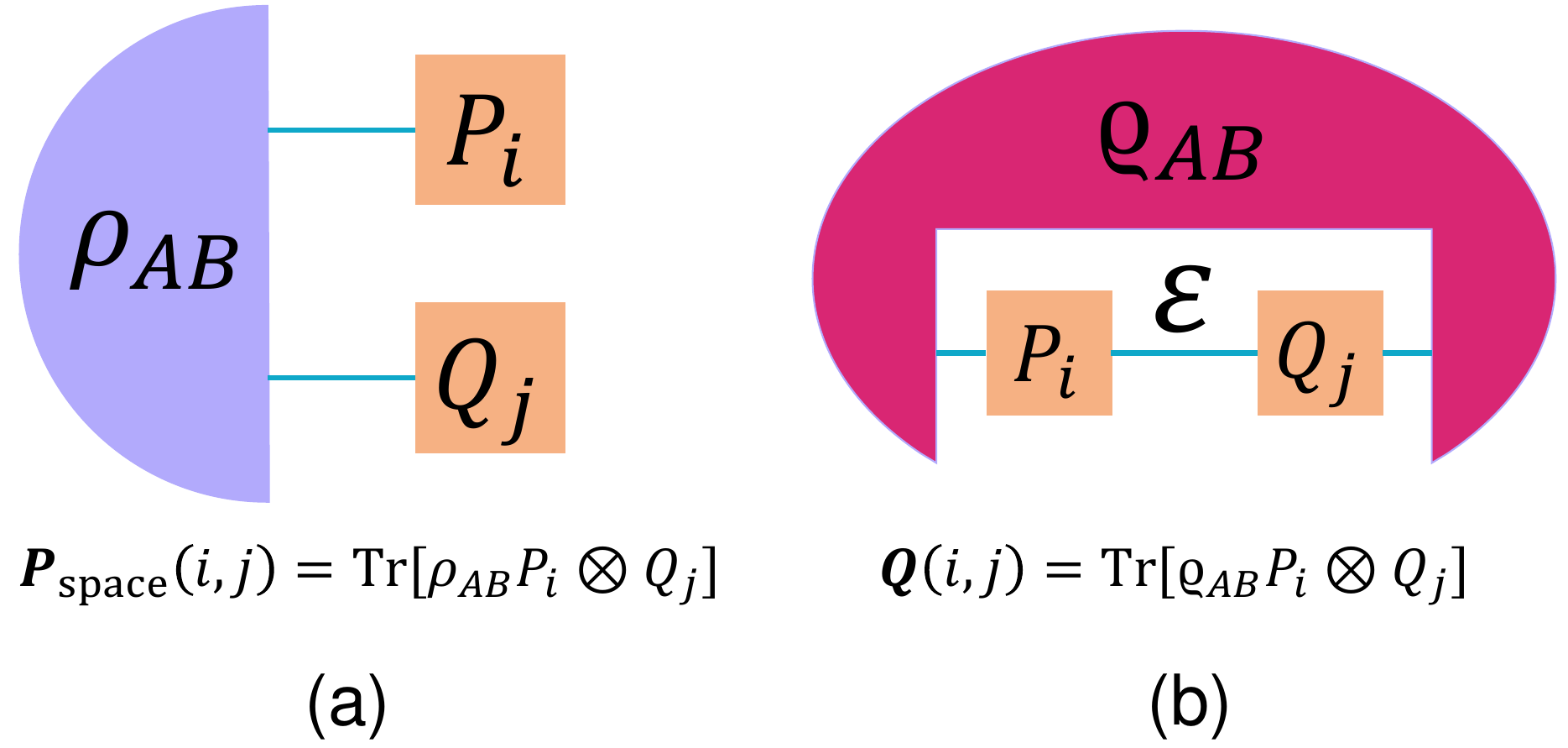}
    \caption{Schematic representation of the Born rules for the distributions $\bold{P}_{\text{space}}$ and $\bold{Q}$. (a) For $\bold{P}_{\text{space}}$, spacelike separated systems $A$ and $B$ which are in a joint spatial state $\rho_{AB}$ are measured in parallel. (b) For $\bold{Q}$, timelike separated systems $A$ and $B$ which are in a joint spatiotemporal state $\varrho_{AB}$ are measured in sequence.}
    \label{fig1}
\end{figure}

The operator $\varrho_{AB}$ instantiating the Born rule \eqref{STXBORN57} is often referred to as a \emph{quantum state over time}~\cite{HHPBS17,FuPa22,FuPa24}, and has been uniquely characterized from various perspectives in Refs.~\cite{LiNg23,PFBC23,FuPa24,FuPa24a,LiFu}. While the operator $\varrho_{AB}$ is Hermitian and of unit trace, it is not positive semi-definite in general, which is a reflection of the fact that the spatiotemporal Born rule \eqref{STXBORN57} is quasiprobabilistic. For systems of qubits, the operator $\varrho_{AB}$ coincides with the \emph{pseudo-density operator} first introduced by Fitzsimons, Jones and Vedral for characterizing quantum correlations which imply causation~\cite{FJV15}, and have since found applications in a number of areas~\cite{HLiu_2025,Marletto_2020,MVVAPGDG21,Pisar19,FuPa22a,ZPTGVF18,song23,Liu_2024,Liu_2025}, including the black hole information problem, temporal teleportation, channel capacity, and quantum causal inference. Moreover, in Refs.~\cite{FuPa23,FZPV24} it was shown how dynamical variants of von~Neumann entropy and quantum mutual information may be extracted from the spectrum of $\varrho_{AB}$, thus laying the foundation for a spatiotemporal formulation of quantum information theory.  

Throughout the course of this work, we also establish several results related to the spatiotemporal Born rule. For clarity of presentation, the remainder of the paper is structured as follows. Section~\ref{S2} introduces the operational framework for two-point sequential measurement scenarios, along with precise definitions of the Treletsky–Margenau–Hill distribution and the modification term $\bold{D}$ in Eq.~\eqref{MHX57}. We then derive key properties of the modification term and explain how it can be experimentally accessed. In Section~\ref{S3}, we present the derivation of the spatiotemporal Born rule \eqref{STXBORN57}. Section~\ref{S4} demonstrates that for sequential measurements of Pauli observables on qubit systems, the temporal correlations encoded in the distributions $\bold{P}_{\text{time}}$ and $\bold{Q}$ are equivalent. From this equivalence, we show that the temporal correlations represented by the quasiprobabilities $\bold{Q}(i,j)$ are directly measurable in such experiments. We also make use of such results to resolve an apparent inconsistency with the pseudo-density matrix formalism. In Section~\ref{S5}, we apply the spatiotemporal Born rule to formulate a new notion of time-reversal symmetry for open quantum systems, illustrating the construction with qubit erasure channels. Finally, Section~\ref{S6} demonstrates that a Born rule of the form \eqref{STXBORN57} for the probabilities $\bold{P}_{\text{time}}$ exists only in the limiting case where $\bold{P}_{\text{time}}(i,j)=\bold{Q}(i,j)$ for all projective measurements $\{P_i\}$ followed by $\{Q_j\}$. As we show that the modification term $\bold{D}$ vanishes precisely when state disturbance from the first measurement does not affect the second---such as when the initial state $\rho_A$ is maximally mixed or when $\E$ acts a discard-and-prepare channel---this finding clarifies how state disturbance acts as an obstruction to the existence of a spatiotemporal Born rule for the distribution $\bold{P}_{\text{time}}$. We conclude with a discussion of our results and potential directions for future research.

\section{Two-point sequential measurement scenarios} \label{S2}

Throughout this work, $A$ and $B$ will denote finite-dimensional quantum systems with Hilbert spaces $\H_A$ and $\H_B$, respectively. The algebra of linear operators on $\H_A$ will be denoted by $\A$ and the algebra of linear operators on $\H_B$ will be denoted by $\B$. The density operators on $\H_A$ and $\H_B$ will be denoted by $\mathfrak{S}(A)$ and $\mathfrak{S}(B)$, respectively. A linear map $\E:\A\to \B$ which is completely positive and trace-preserving (CPTP) will be referred to as a \define{quantum channel}\hspace{0.5mm}.  The Hilbert-Schmidt adjoint of a quantum channel $\E:\A\to \B$ will be denoted by $\E^*:\B\to \A$. A collection of operators $\{P_i\}$ on a finite-dimensional Hilbert space $\H$ is said to be a \define{projective measurement} if and only if $P_iP_j=\delta_{ij}P_i$ and $\sum_iP_i=\mathds{1}$.

\bd
A \define{two-point sequential measurement} consists of the following protocol:
\begin{itemize}
\item
System $A$ is prepared in state $\rho$.
\item
A projective measurement $\{P_i\}$ is then performed on system $A$.
\item
System $A$ then evolves according to a quantum channel $\E:\A\to \B$.
\item
A projective measurement $\{Q_j\}$ is then performed on system $B$.
\end{itemize}
\ed

Such a two-point sequential measurement (TPSM) scenario will be denoted by the 4-tuple $(\rho,\{P_i\},\E,\{Q_j\})$. The \define{L\"{u}ders-von~Neumann (LvN) distribution} associated with $(\rho,\{P_i\},\E,\{Q_j\})$ (which was previously denoted by $\bold{P}_{\text{time}}$) is the probability distribution $\bold{P}$ given by
\be \label{VNDX67}
\bold{P}(i,j)=\Tr[\E(P_i\rho P_i)Q_j]\, ,
\ee
while the \define{Treletsky-Margenau-Hill (TMH) distribution} associated with $(\rho,\{P_i\},\E,\{Q_j\})$ is the quasiprobability distribution $\bold{Q}$ given by
\be \label{MHDX87}
\bold{Q}(i,j)=\frac{1}{2}\Tr[\E(\rho P_i+P_i\rho)Q_j]\, .
\ee
Contrary to the LvN distribution $\bold{P}$, the TMH distribution $\bold{Q}$ is additive with respect to a coarse-graining of projective measurements. In particular, if $P_{i_1},P_{i_2}\in \{P_i\}$ and $P'_{k_0}\in \{P'_k\}$ are such that $P'_{k_0}=P_{i_1}+P_{i_2}$, then for all projective measurements $\{Q_j\}$ we have $\bold{Q}(k_0,j)=\bold{Q}(i_1,j)+\bold{Q}(i_2,j)$, which is a crucial property that distinguishes $\bold{Q}$ from $\bold{P}$. 

The following proposition gives a formula for the TMH distribution $\bold{Q}$ in terms of the LvN distribution $\bold{P}$, and also provides several sufficient conditions for when $\bold{Q}(i,j)=\bold{P}(i,j)$. 

\bn \label{PRXS1}
Let $(\rho,\{P_i\},\E,\{Q_j\})$ be a TPSM scenario, and let $\bold{D}$ be the function given by
\be \label{DMX57}
\bold{D}(i,j)=\frac{1}{2}\Tr[\E(\rho-\rho_i)Q_j]\, , 
\ee
where $\rho_i=P_i\rho P_i+(\mathds{1}-P_i)\rho(\mathds{1}-P_i)$. Then the following statements hold.
\begin{enumerate}[i.] 
\item \label{PRXS11}
$\bold{Q}(i,j)=\bold{P}(i,j)+\bold{D}(i,j)$ for all $i$ and $j$.
\item  \label{PRXS117}
If $\rho$ is the maximally mixed state, then $\bold{D}(i,j)=0$ for all $i$ and $j$. 
\item  \label{PRXS119}
If $\E$ is a discard-and-prepare channel, i.e., if there exists $\sigma\in \S(B)$ such that $\E(a)=\Tr[a]\sigma$ for all $a\in \A$, then $\bold{D}(i,j)=0$ for all $i$ and $j$. 
\item \label{PRXS12}
If $[\rho, P_i]=0$ then $\bold{D}(i,j)=0$ for all $j$. 
\item \label{PRXS13}
If $[P_i,\E^*(Q_j)]=0$ then $\bold{D}(i,j)=0$. 
\end{enumerate}
\en

\bprf
Indeed, for all $i$ and $j$ we have
\begin{align*}
\bold{D}(i,j)&=\frac{1}{2}\Tr[\E(\rho -\rho_i)Q_j] \\
&=\frac{1}{2}\Tr\Big[\E\left(\rho-P_i\rho P_i-(\mathds{1}-P_i)\rho(\mathds{1}-P_i)\right)Q_j\Big] \\
&=\frac{1}{2}\Tr\Big[\E(\rho-2P_i\rho P_i-\rho+\{\rho\, ,P_i\})Q_j\Big] \\
&=-\Tr\Big[\E(P_i \rho P_i) Q_j\Big]+\frac{1}{2}\Tr\Big[\E(\{\rho,P_i\})Q_j\Big] \\
&=-\bold{P}(i,j)+\bold{Q}(i,j)\, ,
\end{align*}
hence $\bold{Q}(i,j)=\bold{P}(i,j)+\bold{D}(i,j)$, thus proving item~\ref{PRXS11}. The proofs of items~\ref{PRXS117}-\ref{PRXS13} are similar. 
\eprf

We note that the modification term $\bold{D}(i,j)$ is measurable for all $i$ and $j$, as it may be written as
\[
\bold{D}(i,j)=\frac{1}{2}\Big(\<Q_j\>_{\E(\rho_i)}-\<Q_j\>_{\E(\rho)}\Big)\, ,
\]  
where $\<\O\>_{\sigma}$ is general notation for the expectation value of an observable $\O$ on a system prepared in state $\sigma$. As such, while the TMH distribution is defined in terms of the same data as that of the LvN distribution---namely, that of a 4-tuple $(\rho,\{P_i\},\E,\{Q_j\})$ uniquely determining a TPSM scenario---the operational realization of the quasiprobabilities $\bold{Q}(i,j)$ requires more than just repeated sequential measurements of $\{P_i\}$ followed by $\{Q_j\}$.  In particular, another protocol is needed to measure the modification term $\bold{D}$. In Section~\ref{S3}, we show how the quasiprobabilities are experimentally accessible via a pair of quantum operations $(\E^{+},\E^{-})$ (associated with the channel $\E$) together with classical post-processing. In Section~\ref{S4}, we show that although the quasiprobabilities $\bold{Q}(i,j)$ are not directly observable from sequential measurements, the temporal \emph{correlations} associated with the quasiprobabilities $\bold{Q}(i,j)$ are indeed directly observable from the operational setup of a TPSM scenario $(\rho,\{P_i\},\E,\{Q_j\})$.

\bx
Consider the TPSM scenario $(\rho,\{P_1,P_2\},\E,\{Q_1,Q_2\})$, where $\rho=|0\>\<0|$ is the ground state of a single qubit, $\E$ the identity channel, $\{P_1,P_2\}=\{\Pi_x^+,\Pi_x^-\}$ and $\{Q_1,Q_2\}=\{\Pi_z^+,\Pi_z^-\}$, where $\Pi_x^{\pm}$ and $\Pi_z^{\pm}$ are the projectors onto the $\pm 1$ eigenspaces of the Pauli spin matrices $\sigma_x$ and $\sigma_z$, respectively. It then follows that the LvN distribution is uniform, so that $\bold{P}(i,j)=1/4$ for all $i,j\in \{1,2\}$. If we take a coarse-graining of the first measurement $\{P_1,P_2\}$ to form the trivial measurement $\{P_0=P_1+P_2=\mathds{1}\}$, then 
\begin{align*}
\bold{P}(0,1)&=\Tr[(\Pi_x^++\Pi_x^-)\rho (\Pi_x^++\Pi_x^-) \Pi_z^+] \\
&=\Tr[\rho \Pi_z^+]=1\, .
\end{align*}
On the other hand, we have
\begin{align*}
\bold{P}(1,1)+\bold{P}(2,1)&=\Tr[\Pi_x^+\rho\Pi_x^+ \Pi_z^+]+\Tr[\Pi_x^-\rho\Pi_x^- \Pi_z^+] \\
&=\frac{1}{4}+\frac{1}{4}=\frac{1}{2}\neq\bold{P}(0,1)\, ,
\end{align*} 
thus the LvN distribution $\bold{P}$ is not locally additive with respect to a coarse-graining of the first measurement. As we show later on, this is due to the fact that the modification term $\bold{D}$ has non-zero values. In particular, we have
$\bold{D}(1,1)=\bold{D}(2,1)=1/4$ and $\bold{D}(1,2)=\bold{D}(2,2)=-1/4$, which is a measure of how state-disturbance due to the outcome of the first measurement affects the likelihood of the second measurement.

The associated TMH distribution in such a context is the genuine probability distribution given by $\bold{Q}(1,1)=\bold{Q}(2,1)=1/2$ and $\bold{Q}(1,2)=\bold{Q}(2,2)=0$. Interestingly, such a distribution $\bold{Q}$ is precisely the distribution one would obtain by abandoning the L\"{u}ders-von Neumann projection postulate, or rather, by assuming that the initial state $\rho=|0\>\<0|$ is undisturbed by the initial measurement. 
\ex

We conclude this section with a result which identifies a class of TPSM scenarios for which the LvN and the TMH distributions in fact coincide.

\bn
Let $\mathcal{I}:\A\to \A$ be the identity channel, and let $\{P_i\}$ be a projective measurement on system $A$. Then for every TPSM scenario of the form $(\rho,\{P_i\},\mathcal{I},\{P_j\})$, we have $\bold{D}(i,j)=0$ for all $i$ and $j$. Equivalently, we have $\bold{P}(i,j)=\bold{Q}(i,j)$ for all $i$ and $j$.
\en

\bprf
Fix $i$ and $j$, and note that
\begin{align*}
\rho_i&=P_i\rho P_i+(\mathds{1}-P_i)\rho(\mathds{1}-P_i) \\
&=2P_i\rho P_i+\rho-\{\rho,P_i\}\, .
\end{align*}
If $i\neq j$, we then have
\[
(\rho-\rho_i)P_j=(\{\rho,P_i\}-2P_i\rho P_i)P_j=P_i\rho P_j\, ,
\]
thus
\[
\bold{D}(i,j)=\frac{1}{2}\Tr[(\rho-\rho_i)P_j]=\frac{1}{2}\Tr[P_i\rho P_j]=0\, .
\]
On the other hand, if $i=j$, we have
\begin{align*}
(\rho-\rho_i)P_j&=(\{\rho,P_i\}-2P_i\rho P_i)P_i \\
&=\rho P_i +P_i\rho P_i-2P_i\rho P_i \\
&=\rho P_i-P_i\rho P_i
\end{align*}
thus
\[
\bold{D}(i,j)=\frac{1}{2}\Tr[(\rho-\rho_i)P_j]=\frac{1}{2}\Tr[\rho P_i-P_i\rho P_i]=0\, ,
\]
as desired.
\eprf

 \section{The spatiotemporal Born rule} \label{S3}

In this section, we prove that given a fixed initial state $\rho_A$ and a quantum channel $\E$ governing the dynamics between measurements associated with a TPSM scenario, then there exists a unique operator $\varrho_{AB}$ encoding the quasiprobabilities $\bold{Q}(i,j)$ as $\{P_i\}$ and $\{Q_j\}$ vary across all projective measurements, thus extending the spacelike Born rule \eqref{PAXRQ87} into the temporal domain.
\bt \label{MTX1}
Fix a quantum channel $\E:\A\to \B$ and a state $\rho\in \S(A)$. Then there exists a unique operator $\varrho_{AB}\in \A\otimes \B$ such that for all TPSM scenarios $(\rho,\{P_i\},\E,\{Q_j\})$,
\be \label{STXBR571}
\bold{Q}(i,j)=\Tr[\varrho_{AB} (P_i\otimes Q_j)]\, ,
\ee
where $\bold{Q}$ is the TMH distribution as defined by \eqref{MHDX87}. Moreover, the operator $\varrho_{AB}$ is given by
\be \label{QSOT91}
\varrho_{AB}=\frac{1}{2}\Big\{\rho\otimes \mathds{1}\, ,\J[\E]\Big\}\, ,
\ee
where $\{\cdot,\cdot\}$ denotes the anticommutator and  $\J[\E]=\sum_{i,j}|i\>\<j|\otimes\E(|j\>\<i|)$ is the Jamio\l kowski operator associated with $\E:\A\to \B$~\cite{Ja72}.
\et

\bprf
First, we show that Eq.~\eqref{STXBR571} holds with $\varrho_{AB}$ as given by \eqref{QSOT91}. For this, let $\{P_i\}$ and $\{Q_j\}$ be projective measurements on $A$ and $B$ respectively. Then for all $i$ and $j$ we have
\begin{align*}
\Tr[\varrho_{AB} (P_i\otimes Q_j)]&=\Tr\left[\frac{1}{2}\Big\{\rho\otimes \mathds{1}\, ,\J[\E]\Big\}(P_i\otimes Q_j)\right] \\
&=\frac{1}{2}\Tr\Big[\J[\E](\{\rho,P_i\}\otimes Q_j)\Big] \\
&=\frac{1}{2}\Tr\Big[\E(\{\rho,P_i\})Q_j\Big]=\bold{Q}(i,j)\, ,
\end{align*}
where the third equality follows from the fact that $\Tr[\J[\E](a\otimes b)]=\Tr[\E(a)b]$ for all operators $a\in \A$ and $b\in \B$. To show that $\varrho_{AB}$ is the unique operator satisfying \eqref{STXBR571}, suppose $\varrho'_{AB}$ is an operator possibly different from $\varrho_{AB}$ such that $\Tr[\varrho'_{AB} (P_i\otimes Q_j)]=\Tr[\varrho_{AB}(P_i\otimes Q_j)]$ for all projectors $P_i$ on $A$ and $Q_j$ on $B$. It then follows from the spectral theorem together with the linearity of trace that $\Tr[XY]=0$ for all observables $Y$ on $AB$, where $X=\varrho'_{AB}-\varrho_{AB}$. The non-degeneracy of the Hilbert-Schmidt inner product then implies $\varrho'_{AB}=\varrho_{AB}$, as desired.
\eprf

Equation~\eqref{STXBR571} will be referred to as the \define{spatiotemporal Born rule} associated with the quantum channel $\E:\A\to \B$ and the initial state $\rho\in \S(A)$. Although the spatiotemporal Born rule is defined for sequential projective measurements, formula  \eqref{STXBR571} naturally extends to positive operator-valued measures (POVMs) by replacing $P_i$ and $Q_j$ with POVM elements $M_i$ and $N_j$. We note that a consequence of the spatiotemporal Born rule \eqref{STXBR571} is that the quasi-probabilities define a bi-additive measure on the space of separable projectors, in accordance with the probabilities $\bold{P}_{\text{space}}(i,j)$ associated with spacelike separated measurements. 

\begin{remark}
The operator $\varrho_{AB}$ given by \eqref{QSOT91} may be decomposed as 
\[
\varrho_{AB}=(\mathcal{I}_A\otimes \E)\big(\mathcal{B}(\rho)\big)\, ,
\]
where $\mathcal{I}_A:\A\to \A$ is the identity channel on system $A$ and $\mathcal{B}:\A\to \A\otimes \A$ is the canonical \emph{virtual broadcasting map}~\cite{PFBC23}. Denoting the swap operator on two copies of system $A$ by $S:\H_A\otimes \H_A\to \H_A\otimes \H_A$, the virtual broadcasting map $\mathcal{B}$ is the hermitian-preserving trace-preserving (HPTP) map given by 
\[
\mathcal{B}(\rho)=\frac{1}{2}\big\{\rho\otimes \mathds{1}, S\big\}\qquad \forall \rho\in \S(A)\, .
\]
In Ref.~\cite{PFBC23}, it was shown that the virtual broadcasting map $\mathcal{B}$ may be decomposed as 
\[
\mathcal{B}=\frac{d+1}{2}\mathcal{B}^+-\frac{d-1}{2}\mathcal{B}^-\, ,
\]
where $d=\text{dim}(\H_A)$, $\mathcal{B}^{\pm}$ are the CPTP maps given by
\[
\mathcal{B}^{\pm}(\rho)=\frac{2}{d\pm 1}\Pi^{\pm}(\mathds{1}\otimes \rho)\Pi^{\pm}\quad \forall \rho\in \S(A)\, ,
\]
and where $\Pi^{\pm}=(\mathds{1}\otimes \mathds{1}\pm S)/2$. It then follows that the operator $\varrho_{AB}$ may be given by the formula
\[
\varrho_{AB}=\frac{d+1}{2}\E^{+}(\rho)-\frac{d-1}{2}\E^{-}(\rho)\, ,
\]
where $\E^{\pm}=(\mathcal{I}_A\otimes \E)\circ \mathcal{B}^{\pm}$. As the maps $\E^{\pm}$ are both CPTP, it follows from the spatiotemporal Born Rule \eqref{STXBR571} that for every TPSM scenario $(\rho,\{P_i\},\E,\{Q_j\})$, the quasiprobabilities $\bold{Q}(i,j)$ are experimentally accessible via the quantum operations $\E^{\pm}$ together with a classical post-processing of the experimental data~\cite{BDOV13}.  
\end{remark}

\section{2-time correlators for systems of qubits} \label{S4}

In this section, we show that in the context of sequential measurements of Pauli observables on systems of qubits, the temporal correlations encoded by the LvN and TMH distributions are in fact equivalent. Remarkably, such a result implies that in such a context the temporal correlations associated with TPSM scenarios are independent of the L\"{u}ders-von Neumann projection postulate. Moreover, a direct consequence of this result is that even though the TMH distribution is not directly observable from sequential measurements, the correlations encoded by the TMH distribution are directly observable from the operational setup of a TPSM scenario. In such a context, the operator $\varrho_{AB}$ instantiating the spatiotemporal Born rule \eqref{STXBR571} is in fact a pseudo-density matrix~\cite{FJV15}, which is a generalization of a multi-qubit density matrix which encodes correlations across both space and time. The equivalence of the temporal correlations encoded by the LvN and TMH distributions together with the spatiotemporal Born rule will then be used clarify a misconception about the pseudo-density matrix formalism. 

So now suppose that system $A$ consists of $m$ qubits and system $B$ consists of $n$ qubits, and consider Pauli observables $\sigma_A=\sigma_{\alpha_1}\otimes \cdots \otimes \sigma_{\alpha_{m}}$  and $\sigma_B=\sigma_{\beta_1}\otimes \cdots \otimes \sigma_{\beta_{n}}$ with $\alpha_i,\beta_j\in \{0,\ldots,3\}$. If $\sigma_A$ and $\sigma_B$ are not the identity matrix, then $\sigma_{A}=\Pi^+-\Pi^-$ and $\sigma_{A}=\Xi^+-\Xi^-$, where $\Pi^{\pm}$ and $\Xi^{\pm}$ are the orthogonal projectors onto the $\pm1$-eigenspaces of $\sigma_A$ and $\sigma_B$, respectively. We then consider a TPSM scenario of the form 
\be \label{TPSX17}
\big(\rho,\{\Pi^{\pm}\},\E,\{\Xi^{\pm}\}\big)\, ,
\ee
so that $\rho$ is a density operator on $\H_A$ and $\E:\A\to \B$ is a quantum channel. In such a case, we define the \define{L\"{u}ders-von Neumann (LvN) 2-time correlator} to be the element $\<\sigma_{A},\sigma_{B}\>_{_{\text{LvN}}}\in [-1,1]$ given by
\[
\<\sigma_{A},\sigma_{B}\>_{_{\text{LvN}}}=\bold{P}(+,+)-\bold{P}(+,-)-\bold{P}(-,+)+\bold{P}(-,-)\, ,
\]
where $\bold{P}(\pm,\pm)$ is the LvN distribution associated with the TPSM scenario \eqref{TPSX17}. Similarly, we define the \define{Treletsky-Margenau-Hill (TMH) 2-time correlator} to be the element $\<\sigma_{A},\sigma_{B}\>_{_{\text{TMH}}}\in [-1,1]$ given by
\[
\<\sigma_{A},\sigma_{B}\>_{_{\text{TMH}}}=\bold{Q}(+,+)-\bold{Q}(+,-)-\bold{Q}(-,+)+\bold{Q}(-,-)\, ,
\]
where $\bold{Q}(\pm,\pm)$ is the TMH distribution associated with the TPSM scenario \eqref{TPSX17}. It follows immediately from their definitions that the two-time correlators $\<\sigma_{A},\sigma_{B}\>_{_{\text{LvN}}}$ and $\<\sigma_{A},\sigma_{B}\>_{_{\text{TMH}}}$ correspond to the weighted averages of the \emph{product} of measurements $\{\Pi^{\pm}\}$ followed by $\{\Xi^{\pm}\}$ with respect to the LvN and TMH distributions, respectively. Moreover, the definitions of $\<\sigma_{A},\sigma_{B}\>_{_{\text{LvN}}}$ and $\<\sigma_{A},\sigma_{B}\>_{_{\text{TMH}}}$ easily extend to the case when either $\sigma_A$ or $\sigma_B$ is the identity matrix by considering the associated projective measurement to be the trivial projective measurement $\{\mathds{1}\}$. 

We now prove the following:

\bt \label{TXM27}
Under the assumptions above, we have
\[
\<\sigma_{A},\sigma_{B}\>_{_{\emph{LvN}}}=\<\sigma_{A},\sigma_{B}\>_{_{\emph{TMH}}}=\Tr[\varrho_{AB} (\sigma_A\otimes \sigma_B)]\, ,
\]
where $\varrho_{AB}=\frac{1}{2}\big\{\rho\otimes \mathds{1},\mathscr{J}[\E]\big\}$. 
\et

\bprf
We first consider the case when $\sigma_A$ and $\sigma_B$ are both not the identity operator. In such a case we have $\bold{Q}(\pm,\pm)=\bold{P}(\pm,\pm)+\bold{D}(\pm,\pm)$, thus to prove $\<\sigma_{A},\sigma_{B}\>_{_{\text{LvN}}}=\<\sigma_{A},\sigma_{B}\>_{_{\text{TMH}}}$ it suffices to show
\be \label{D07}
\bold{D}(+,+)-\bold{D}(+,-)-\bold{D}(-,+)+\bold{D}(-,-)=0\, .
\ee
Indeed, by the definition of $\bold{D}$ as given by \eqref{DMX57}, we have
\[
\bold{D}(\pm,\pm)=\frac{1}{2}\Tr\big[\E(\rho-\rho^{\pm})\Xi^{\pm}\big]\, ,
\]
where $\rho^{\pm}=\Pi^{\pm}\rho\Pi^{\pm}+(\mathds{1}-\Pi^{\pm})\rho(\mathds{1}-\Pi^{\pm})$. Now since $\Pi^++\Pi^-=\mathds{1}$, it follows that $\rho^+=\rho^-$. Therefore, we have
\[
\bold{D}(+,+)-\bold{D}(-,+)=\frac{1}{2}\Tr\big[\E(\rho^--\rho^{+})\Xi^{+}\big]=0\, ,
\]
and 
\[
\bold{D}(+,+)-\bold{D}(-,+)=\frac{1}{2}\Tr\big[\E(\rho^{+}-\rho^{-})\Xi^{-}\big]=0\, ,
\]
thus \eqref{D07} indeed holds. Moreover, since 
\begin{align*}
\sigma_A\otimes \sigma_B&=\Pi^+\otimes \Xi^+-\Pi^+\otimes \Xi^- \\
&-\Pi^-\otimes \Xi^++\Pi^-\otimes\Xi^-\, ,
\end{align*}
the equality $\<\sigma_{A},\sigma_{B}\>_{_{\text{TMH}}}=\Tr[\varrho_{AB} (\sigma_A\otimes \sigma_B)]$ follows from the spatiotemporal Born rule \eqref{STXBR571} and the linearity of trace. The case that either $\sigma_A$ or $\sigma_B$ is the identity operator follows from similar calculations, thus concluding the proof.
\eprf

We note that it follows from Theorem~\ref{TXM27} that not only are the correlations encoded by the LvN and TMH distributions equivalent in such a context, but such correlations are also encoded by the operator $\varrho_{AB}$ instantiating the spatiotemporal Born rule \eqref{STXBR571}. Moreover, the fact that $\<\sigma_{A},\sigma_{B}\>_{_{\text{LvN}}}=\<\sigma_{A},\sigma_{B}\>_{_{\text{TMH}}}$ also implies that the correlations associated with the TMH distribution are directly observable by the operational setup of the TPSM scenario \eqref{TPSX17}. In particular, if the protocol corresponding to the TPSM scenario \eqref{TPSX17} is repeated many times and one collects the data corresponding to the product of the of the first and second measurements for each run of the experiment, then the average of such products should converge to $\<\sigma_{A},\sigma_{B}\>_{_{\text{TMH}}}$ as the number of repetitions becomes large. 

If we denote the Pauli basis of observables on $A$ by $\{\sigma_{\alpha}\}$ and the Pauli basis of observables on $B$ by $\{\sigma_{\beta}\}$ (so that $\alpha$ and $\beta$ are vectorial indices), then it follows from Theorem~\ref{TXM27} together with properties of Pauli matrices that
\be \label{PDMXS71}
\varrho_{AB}=\frac{1}{2^{m+n}}\sum_{\alpha,\beta}\<\sigma_{\alpha},\sigma_{\beta}\>_{_{\text{LvN}}}\sigma_{\alpha}\otimes \sigma_{\beta}\, ,
\ee
where $\varrho_{AB}=\frac{1}{2}\{\rho\otimes \mathds{1},\mathscr{J}[\E]\}$ is the operator instantiating the spatiotemporal Born rule \eqref{STXBR571}. The RHS of Eq.~\eqref{PDMXS71} is the definition of the \emph{pseudo-density matrix} associated with the quantum channel $\E:\A\to \B$ and the state $\rho\in \S(A)$. It then follows that in such a context, the operator $\varrho_{AB}$ is precisely the pseudo-density matrix associated with the pair $(\rho,\E)$. We note that if the 2-time expectation values $\<\sigma_{\alpha},\sigma_{\beta}\>_{_{\text{LvN}}}$ appearing on the RHS of Eq.~\eqref{PDMXS71} are replaced by the spacelike expectation values $\<\sigma_{\alpha}\otimes \sigma_{\beta}\>\equiv \Tr[\rho_{AB}(\sigma_{\alpha}\otimes \sigma_{\beta})]$ for some density matrix $\rho_{AB}$, then one obtains the density matrix $\rho_{AB}$. As such, the pseudo-density matrix $\varrho_{AB}$ encodes temporal correlations in precisely the same manner that a density matrix $\rho_{AB}$ encodes spatial correlations. 

Although the pseudo-density matrix construction provides a simple extension of the density matrix formalism into the temporal domain, a viewpoint held by many is that there is an inconsistency with the notion of a pseudo-density matrix $\varrho_{AB}$ as given by \eqref{PDMXS71}. In particular, while it is always the case that
\[
\Tr[\varrho_{AB}(\sigma_{A}\otimes \sigma_{B})]=\<\sigma_{A},\sigma_{B}\>_{_{\text{LvN}}}\quad \forall \sigma_{A},\sigma_{B}\, ,
\]
in general we have that if $\sigma_{A}=\Pi^+-\Pi^-$ and $\sigma_{B}=\Xi^+-\Xi^-$, then
\[
\Tr[\varrho_{AB}(\Pi^{\pm}\otimes \Xi^{\pm})]\neq \bold{P}(\pm,\pm)\, .
\]
In other words, while pseudo-density matrices encode the correlations $\<\sigma_{A},\sigma_{B}\>_{_{\text{LvN}}}$, they do not encode the associated probabilities $\bold{P}(\pm,\pm)$ corresponding to the LvN distribution. However, in light of Theorem~\ref{TXM27}, we have
\[
\Tr[\varrho_{AB}(\sigma_{A}\otimes \sigma_{B})]=\<\sigma_{A},\sigma_{B}\>_{_{\text{TMH}}}\quad \forall \sigma_{A},\sigma_{B}\, ,
\]
and by the spatiotemporal Born rule \eqref{STXBR571} it follows that if $\sigma_{A}=\Pi^+-\Pi^-$ and $\sigma_{B}=\Xi^+-\Xi^-$, then
\[
\Tr[\varrho_{AB}(\Pi^{\pm}\otimes \Xi^{\pm})]= \bold{Q}(\pm,\pm)\, .
\]
As such, the apparent inconsistency of the pseuso-density matrix construction stems from viewing its associated correlations as $\<\sigma_{A},\sigma_{B}\>_{_{\text{LvN}}}$ rather than $\<\sigma_{A},\sigma_{B}\>_{_{\text{TMH}}}$. It then follows that once the correlations associated with a pseudo-density matrix are identified with $\<\sigma_{A},\sigma_{B}\>_{_{\text{TMH}}}$, there is no longer any inconsistency between the correlations and the (quasi)probabilities encoded by a pseudo-density matrix.

\section{Time-reversal symmetry and a spatiotemporal Bayes' rule} \label{S5}

Given a pair of classical random variables $(X,Y)$ measured in sequence, the associated joint distribution $\mathbb{P}(x,y)$ is symmetric in its arguments. This is due to the fact that the joint probability that $X=x$ and $Y=y$ is invariant under a reversal of the order in which $X$ and $Y$ were measured, which is a characteristic feature of classical random variables. At the quantum level however, state collapse combined with noise due to system-environment interactions yields an apparent time \emph{asymmetry} associated with sequential measurements performed on a quantum system. In this section, we show how the spatiotemporal Born rule \eqref{STXBR571} combined in conjunction with the quantum Bayes' rule from \cite{FuPa22a} restores time-reversal symmetry for the TMH distribution associated with TPSM scenarios, even in the presence of noise.

For this, let $\F:\B\to \A$ be a quantum channel, and let $\varrho_{BA}$ be the bipartite operator on $\H_B\otimes \H_A$ given by
\[
\varrho_{BA}=\frac{1}{2}\Big\{\E(\rho)\otimes \mathds{1}\, ,\J[\F]\Big\}\, ,
\]
where $\E$ and $\rho$ are fixed as in the spatiotemporal Born rule \eqref{STXBR571}. The operator $\varrho_{BA}$ is then said to satisfy the \emph{quantum Bayes' rule} if and only if $S\varrho_{BA}S=\varrho_{AB}$, where $S=\sum_{i,j}|i\>\<j|\otimes|j\>\<i|$ is the swap operator and $\varrho_{AB}$ is given by \eqref{QSOT91}. In such a case the channel $\F$ is then said to be a \emph{Bayesian inverse} of $\E$ with respect to $\rho$. Now let $\overline{Q}$ be the TMH distribution associated with TPSM scenarios of the form $(\E(\rho),\{Q_j\},\F,\{P_i\})$. For all $i$ and $j$ we then have
\begin{align*}
\overline{\bold{Q}}(j,i)&=\Tr[\varrho_{BA} (Q_j\otimes P_i)]=\Tr[S\varrho_{BA}(Q_j\otimes P_i)S] \\
&=\Tr[S\varrho_{BA}SS(Q_j\otimes P_i)S] \\
&=\Tr[\varrho_{AB}(P_i\otimes Q_j)]=\bold{Q}(i,j)\, ,
\end{align*}
where the first and final equalities follow from the spatiotemporal Born rule, and the fourth equality follows from the quantum Bayes' rule $S\varrho_{BA}S=\varrho_{AB}$. As such, when $\varrho_{BA}$ satisfies the quantum Bayes' rule, the TPSM scenario $(\E(\rho),\{Q_j\},\F,\{P_i\})$ may be viewed as an operational time-reversal of the TPSM scenario $(\rho,\{P_i\},\E,\{Q_j\})$, thus restoring the symmetry $\bold{Q}(i,j)=\overline{\bold{Q}}(j,i)$ of classical joint probability distributions. After multiplying the LHS and RHS of the equation $\bold{Q}(i,j)=\overline{\bold{Q}}(j,i)$ by $\bold{Q}(i)/\bold{Q}(i)$ and $\overline{\bold{Q}}(j)/\overline{\bold{Q}}(j)$ respectively, we then arrive at the equation
\be \label{STBRXS17}
\bold{Q}(j|i)\bold{Q}(i)=\overline{\bold{Q}}(i|j)\overline{\bold{Q}}(j)\, ,
\ee
which we refer to as the \emph{spatiotemporal Bayes' rule} for the TMH distribution.

\bx[Qubit Erasure Channels]
Let $A$ and $B$ be 2- and 3-level systems respectively, with $\H_A\subset \H_B$ a subspace, and let $\E:\A\to \B$ be the \emph{qubit erasure channel}, which is the map given by
\[
\E(\omega)=(1-\lambda)\omega+\lambda\Tr[\omega]|2\rangle \langle 2|\, , \qquad \lambda\in (0,1)\, . 
\]
Given a fixed input state of the form $\rho=p |0\rangle \langle 0|+(1-p)|1\rangle \langle 1|$, the Sylvester equation 
\[
(\E(\rho)\otimes \mathds{1})X+X(\E(\rho)\otimes \mathds{1})=2S\varrho_{AB}S
\]
admits a unique solution $X$, from which it follows that $\F=\J^{-1}[X]$ is the Bayesian inverse of the qubit erasure channel $\E$ with respect to $\rho$. We then find that the map $\F:\B\to \A$ is such that $\F(a)=a$ for all $a\in \A\subset \B$, and $\F(|2\rangle \langle 2|)=\rho$. In such a case, we have $\varrho_{BA}=\sum_{k,l=0}^{1}a_{kl}|k\rangle \langle l|\otimes |l\rangle \langle k|+\lambda |2\rangle \langle 2|\otimes \rho$, where $a_{kl}=(1-\lambda)(a_k+a_l)/2$, with $a_0=p$ and $a_1=(1-p)$, and since $\varrho_{AB}=\sum_{k,l=0}^{1}a_{kl}|k\rangle \langle l|\otimes |l\rangle \langle k| +\lambda \rho\otimes |2\rangle \langle 2|$, it immediately follows that the quantum Bayes' rule $S\varrho_{BA}S=\varrho_{AB}$ is indeed satisfied. It then follows from the spatiotemporal Born rule that $\bold{Q}(i,j)=\overline{\bold{Q}}(j,i)$ for all TPSM scenarios $(\rho,\{P_i\},\E,\{Q_j\})$ and $(\E(\rho),\{Q_j\},\F,\{P_i\})$, thus implying that the spatiotemporal Bayes' rule \eqref{STBRXS17} holds for such an $\E$ and $\F$. Such a time-reversal symmetry is quite striking in light of the fact that in the forward process $\E$, the information of the input state is lost with probability $\lambda$, thus the process $\E$ is typically not viewed as being reversible in any sense.
\ex

In the context of TPSM scenarios $(\rho,\{\Pi^{\pm}\},\E,\{\Xi^{\pm}\})$ associated with sequential measurements of Pauli observables $\sigma_A=\Pi^+-\Pi^-$ and $\sigma_B=\Xi^+-\Xi^-$ on a system of qubits (as considered in Section~\ref{S4}), Theorem~\ref{TXM27} together with the Quantum Bayes' rule $S\varrho_{BA}S=\varrho_{AB}$ implies 
\be \label{TRSX17}
\<\sigma_A,\sigma_B\>_{_{\text{LvN}}}=\<\sigma_B,\sigma_A\>_{_{\text{LvN}}}\, ,
\ee
where $\<\sigma_A,\sigma_B\>_{_{\text{LvN}}}$ is the LvN 2-time correlator associated with the TPSM scenario $(\rho,\{\Pi^{\pm}\},\E,\{\Xi^{\pm}\})$, and $\<\sigma_B,\sigma_A\>_{_{\text{LvN}}}$ is the LvN 2-time correlator associated with the TPSM scenario $(\E(\rho),\{\Xi^{\pm}\},\F,\{\Pi^{\pm}\})$. As such, the Bayesian inverse $\F$ provides an operational form of time-reversal symmetry for the temporal correlations arising in TPSM scenarios of the form  $(\rho,\{\Pi^{\pm}\},\E,\{\Xi^{\pm}\})$. Notably, these correlations are directly accessible within the operational framework of both the original TPSM scenario $(\rho,\{\Pi^{\pm}\},\E,\{\Xi^{\pm}\})$ and its time-reversed counterpart $(\E(\rho),\{\Xi^{\pm}\},\F,\{\Pi^{\pm}\})$. A concrete illustration of the time-reversal symmetry expressed in Eq.~\eqref{TRSX17} was provided for amplitude damping channels in Ref.~\cite{FuPa24a}.

We note that the Petz map $\F_{_{\text{Petz}}}:\B\to \A$ also provides a notion of reversibility for a quantum channel $\E:\A\to \B$ with respect to a reference state $\rho\in \S(A)$. The Petz map is given by
\[
\F_{_{\text{Petz}}}(\sigma)=\sqrt{\rho}\E^*\big(\E(\rho)^{-1/2}\sigma\E(\rho)^{-1/2}\big)\sqrt{\rho}\quad \forall \sigma\in \S(B)\, ,
\]
and is fundamental to the study of information recovery in quantum information theory~\cite{Wilde_22}. Recent work has explored the relationship between the Petz map and the Bayesian inverse. Specifically, for a single-qubit amplitude-damping channel and a diagonal reference state $\rho$, it has been demonstrated that the associated Petz map $\F_{_{\text{Petz}}}$ and the Bayesian inverse $\F$ differ by a dephasing channel~\cite{FuPa24a}. This difference appears to be a general characteristic of the relationship between the two maps across various scenarios~\cite{Song_25}. Consequently, the Petz map $\F_{_{\text{Petz}}}$ will not in general yield an operational time-reversal symmetry for TPSM scenarios as achieved by the Bayesian inverse $\F$. 

In Ref.~\cite{Aw23}, quasiprobability representations of the Petz map $\F_{_{\text{Petz}}}$ are investigated in the context of quantum Bayesian inference. It would be interesting to compare such quasiprobability represetations between the Petz map $\F_{_{\text{Petz}}}$ and the Bayesian inverse $\F$ with respect to the pair $(\rho,\E)$.

\section{Existence of a spatiotemporal Born rule for the L\"{u}ders-von~Neumann distribution} \label{S6}

In this section we prove several necessary and sufficient conditions for the existence of a spatiotemporal Born rule of the form \eqref{STXBR571} for the LvN distribution. As a consequence, we find that such a spatiotemporal Born rule exists if and only if the LvN and the TMH distributions in fact coincide. Since the modification term $\bold{D}$ relating the LvN and TMH distributions vanishes precisely when state disturbance due to the initial measurement has no effect on the outcome of the second measurement, such a result yields a precise explanation for why state disturbance acts as an obstruction to the existence of a Born rule for the LvN distribution. To be absolutely precise about what exactly is meant by the existence of a Born rule for the LvN distribution, fix a quantum channel $\E:\A\to \B$ and an initial state $\rho\in \S(A)$. A spatiotemporal Born rule is then said to exist for the LvN distribution with respect to the pair $(\E,\rho)$ if and only if there exists an operator $\varrho_{AB}$ such that for all TPSM scenarios $(\rho,\{P_i\},\E,\{Q_j\})$,
\be \label{BRXVNL67}
\Tr[\varrho_{AB}(P_i\otimes Q_j)]=\bold{P}(i,j)\equiv \Tr[\E(P_i \rho P_i) Q_j]\, .
\ee 

\bt \label{MTX2}
Fix a quantum channel $\E:\A\to \B$ and a state $\rho\in \S(A)$. Then the following statements are equivalent. 
\begin{enumerate}[i.]
\item \label{MTX21}
The spatiotemporal Born rule exists for the LvN distribution $\bold{P}$ with respect to $(\E,\rho)$.
\item \label{MTX22}
$\bold{P}(i,j)=\bold{Q}(i,j)$ holds for every TPSM scenario $(\rho,\{P_i\},\E,\{Q_j\})$, where $\bold{Q}$ is the TMH distribution \eqref{MHDX87}. 
\item \label{MTX221}
$\bold{D}(i,j)=0$ holds for every TPSM scenario $(\rho,\{P_i\},\E,\{Q_j\})$, where $\bold{Q}$ is the TMH distribution \eqref{MHDX87}. 
\item \label{MTX23}
$\E(\rho-\rho_P)=0$ for all projectors $P$ on $A$, where $\rho_P=P\rho P+(\mathds{1}-P)\rho(\mathds{1}-P)$. 
\item \label{MTX24}
The LvN distribution $\bold{P}$ is additive with respect to a coarse-graining of initial measurements, i.e., if $P_1$ and $P_2$ are orthogonal projectors on $A$ and $P_0=P_1+P_2$, then for all projectors $Q_j$ on $B$,
\[
\bold{P}(0,j)=\bold{P}(1,j)+\bold{P}(2,j)\, .
\]
\end{enumerate}
Finally, when such a spatiotemporal Born rule exists for the LvN distribution $\bold{P}$ with respect to $(\E,\rho)$, the operator $\varrho_{AB}$ instantiating \eqref{BRXVNL67} is unique, and is given by \eqref{QSOT91}.
\et

\bprf
(\ref{MTX21}$\implies$\ref{MTX22}) 
Suppose the spatiotemporal Born rule holds for the L\"{u}ders-von~Neumann distribution, so that there exists an operator $\varrho_{AB}$ such that \eqref{BRXVNL67} holds for every TPSM scenario $(\rho,\{P_i\},\E,\{Q_j\})$. Now let $\{P_1,P_2\}$ be a binary projective measurement on $A$, let $\O_B$ be an observable on $B$, and let $\O_A=P_1-P_2$. In such a case, the two-time expectation value of $\O_A$ followed by $\O_B$ is the real number $\<\O_A,\O_B\>$ given by
\[
\<\O_A,\O_B\>=\Tr[\E(P_1\rho P_1)\O_B]-\Tr[\E(P_2\rho P_2)\O_B]\, ,
\]
and by direct calculation we obtain $\<\O_A,\O_B\>=\Tr[\varrho_{AB} (\O_A\otimes \O_B)]$. Now since $\{P_1,P_2\}$ is an arbitrary binary projective measurement on $A$ and $\O_B$ is an arbitrary observable on $B$, it follows from Theorem~5.4 in Ref.~\cite{FuPa24} that $\varrho_{AB}$ is the operator given by \eqref{QSOT91}, thus by Theorem~\ref{MTX1} and Eq.~\eqref{BRXVNL67} we have $\bold{P}(i,j)=\bold{Q}(i,j)$ for every TPSM scenario $(\rho,\{P_i\},\E,\{Q_j\})$, as desired. 

(\ref{MTX22}$\implies$\ref{MTX221}) The implication follows immediate from item~\ref{PRXS11} of Proposition~\ref{PRXS1}. 

(\ref{MTX221}$\implies$\ref{MTX23}) As $\bold{P}(i,j)=\bold{Q}(i,j)$ holds for every projector $Q_j$, one can obtain $2\E(P_i \rho P_i) = \E(\{P_i,\rho\})$, i.e., \ref{MTX23}.

(\ref{MTX23}$\implies$\ref{MTX24}) Suppose that $P_1 = \sum_{a} |a\rangle\langle a|$ and $P_2 = \sum_b |b\rangle\langle b|$ with $\langle a|b\rangle = 0$ for all $a,b$, the statement (\ref{MTX24}) is equivalent to $\E(P_1\rho P_2 + P_2\rho P_1) = 0$, which can be rewritten as $\sum_{a,b} \langle a|\rho|b\rangle \E(|a\rangle\langle b|) + \langle b|\rho|a\rangle \E(|b\rangle\langle a|)=0$.
 Choosing $P' = |\alpha\rangle\langle \alpha|$, according to $(\ref{MTX23})$, we can obtain that $\E(\{|\alpha\rangle\langle \alpha|,\rho\}) = 2\langle \alpha|\rho|\alpha\rangle\E(|\alpha\rangle\langle \alpha|$ for any pure state $|\alpha\rangle$. Now choose the projector $P = |a\rangle\langle a| + |b\rangle\langle b|$, together with the result about $P'$, (\ref{MTX23}) gives rise to $\langle a|\rho|b\rangle \E(|a\rangle\langle b|) + \langle b|\rho|a\rangle \E(|b\rangle\langle a|) = 0$, which implies (\ref{MTX24}).

 (\ref{MTX24}$\implies$\ref{MTX21}) The statement \ref{MTX24} gives that $\Tr[\E(P_1\rho P_2 + P_2\rho P_1)Q_j ]= 0$ for all projectors $Q_j$ on $B$ and orthogonal projectors $P_1,P_2$ on $A$. By choosing $P_1 = P'_i$ and $P_2 = \mathds{1} - P'_i$, one can get $\bold{P}(i,j) = \bold{Q}(i,j)$ holds for every TPSM scenario $(\rho,\{P'_i\},\E,\{Q_j\})$, where $\bold{Q}$ is the TMH distribution \eqref{MHDX87}. Then by Theorem~\ref{MTX1} we have that \eqref{BRXVNL67} holds with $\varrho_{AB}$ as given by \eqref{QSOT91}, thus the spatiotemporal Born rule holds for the LvN distribution, as desired.
\eprf

As the each of the conditions of Theorem~\ref{MTX2} is highly restrictive, it follows that a Born rule exists for the LvN distribution only for very specialized cases of the dynamics $(\E,\rho)$. In the case that either $\rho$ is maximally mixed or $\E$ is a discard-and-prepare channel---i.e., when there is no back-reaction due to the initial measurement---it follows from items~\ref{PRXS117} and \ref{PRXS119} of Proposition~\ref{PRXS1} that $\bold{P}(i,j)=\bold{Q}(i,j)$ for every TPSM scenario $(\rho,\{P_i\},\E,\{Q_j\})$, thus Theorem~\ref{MTX2} ensures that the spatiotemporal Born rule \eqref{BRXVNL67} exists for the LvN distribution in such a case. It is then natural to ask whether $\rho$ being maximally mixed or $\E$ being a discard-and-prepare channel is also a necessary condition for the existence of the spatiotemporal Born rule for the LvN distribution. While it is straightforward to show that this is indeed the case when $A$ consists of a single qubit, for systems of arbitrary dimension we leave it as an open question.

\section{Discussion}

In this work, we considered the problem of extending the Born rule of quantum theory to the spatiotemporal domain---a task complicated by inherently quantum features such as state collapse and decoherence. The process matrix formalism offers one possible approach to this problem, but it departs formally from standard quantum theory by defining joint probabilities over the broader space of separable instrument components rather than over separable projectors. Here, we adopted a different strategy, formulating a spatiotemporal Born rule that preserves the physical principle of bi-additivity for joint probabilities while relaxing the conventional requirement of non-negativity. Specifically, we demonstrated that, for a fixed initial state and a fixed quantum channel governing dynamics between two sequential measurements, there exists a unique pseudo-density operator that encodes canonical quasiprobabilities in precisely the same manner that a density operator encodes joint probabilities for spacelike separated measurements. This result provides a unified treatment of spatial and temporal correlations in quantum theory without invoking epistemic extensions of the Born rule that rely on higher-order operators to assign joint probabilities over separable instrument components rather than separable projectors.

We also derived various consequences of such a spatiotemporal Born rule. In particular, we showed how to access the quasiprobabilities encoded by the spatiotemporal Born rule by a pair of quantum operations together with classical post-processing. We also showed how such a spatiotemporal Born rule applied in conjunction with the notion of Bayesian inversion for quantum channels yields an operational notion of time-reversal symmetry for open quantum systems, from which one may derive a quasiprobabilistic Bayes' rule for sequential measurements. Such a time-reversal symmetry for open quantum systems is quite striking, as the dynamics of open quantum systems are typically not thought of as being reversible in any sense. 

While we have focused here on the case of bipartite correlations, we note that by employing the multipartite formalism for pseudo-density operators the results in this work readily extend to multi-time sequential measurement scenarios, such as those considered in the context of Leggett-Garg inequalities. A more detailed account of the multi-partite extension of the results established here, and the implications for a realist perspective of the correlations associated with Leggett-Garg scenarios will appear in a later work.

\emph{Acknowledgments}. JF is supported by the Hainan University startup fund for the project ``Spacetime from Quantum Information", and would like to thank Seok Hyung Lie, Xiangjing Liu and Arthur J. Parzygnat for useful discussion. ZM is supported by the the Fundamental Research Funds for the Central Universities,  and NSFC 12371132.







\bibliography{references}

\end{document}